# Non-equilibrium Fluctuational Quantum Electrodynamics: Heat Radiation, Heat Transfer, and Force

G. Bimonte,[1] T. Emig,[2,3] M. Kardar,[4] and M. Krüger[5]

[1]Dipartimento di Fisica E. Pancini, Università di Napoli Federico II, Complesso Universitario MSA, Via Cintia, I-80126 Napoli, Italy and INFN Sezione di Napoli, I-80126 Napoli, Italy
[2]Multi-Scale Materials Science for Energy and Environment, UMI 3466, The joint CNRS-MIT Laboratory, Massachusetts Institute of Technology, Cambridge, Massachusetts 02139, USA
[3]LPTMS, CNRS, Université Paris-Saclay, 91405 Orsay, France
[4]Massachusetts Institute of Technology, Department of Physics, Cambridge, Massachusetts 02139, USA
[5]4th Institute for Theoretical Physics, Universität Stuttgart, Germany and Max Planck Institute for Intelligent Systems, 70569 Stuttgart, Germany

**Abstract**

Quantum and thermal fluctuations of electromagnetic waves are the cornerstone of quantum and statistical physics, and inherent to such phenomena as thermal radiation and van der Waals forces. While the basic principles are the material of elementary texts, recent experimental and technological advances have made it necessary to come to terms with counterintuitive consequences of electromagnetic fluctuations at short scales– in the so called *near-field* regime. We focus on three manifestations of such behavior: **(i)** The Stefan–Boltzmann law describes thermal radiation from macroscopic bodies, but fails to account for magnitude, polarization and coherence of radiation from small objects (say compared to the skin depth). **(ii)** The heat transfer between two bodies at similar close proximity is dominated by evanescent waves, and can be several orders of magnitude larger than the classical contribution due to propagating waves. **(iii)** Casimir/van der Waals interactions are a dominant force between objects at sub-micron separation; the non-equilibrium analogs of this force (for objects at different temperatures) have not been sufficiently explored (at least experimentally). To explore these phenomena we introduce the tool of fluctuational quantum electrodynamics (QED) originally introduced by Rytov in the 1950s. Combined with a scattering formalism, this enables studies of heat radiation and transfer, equilibrium and non-equilibrium forces for objects of different material properties, shapes, separations and arrangements.

# Contents

## 1. INTRODUCTION

The beginnings of quantum mechanics can be traced back to the introduction of Planck's constant $\hbar = h/2\pi$ for describing the spectrum of black body radiation at a given temperature $T$ (1). Planck's law then leads the well-known Stefan-Boltzmann law (2) for radiated power ($\propto \sigma T^4$, where $\sigma$ is the Stefan-Boltzmann constant), and in turn to radiative heat transfer between objects at different temperatures ($\propto \sigma(T_1^4 - T_2^4)$. To first approximation, everyday matter is held together by the fluctuating electromagnetic fields between (on average) neutral objects. At the atomic scale, this attractive interaction appears in the guises of van der Waals, Keesom, Debye, and London forces (3). At larger scales, the collective behavior of condensed atoms is better formulated in terms of dielectric properties. In 1948, Casimir computed the force between two perfectly conducting parallel plates which also arises due to the quantum fluctuations of the electromagnetic waves in the intervening vacuum (or, equivalently, due to the charge and current fluctuations in the plates) (4). This calculation was extended by Lifshitz to dielectric plates, accounting for the fluctuating currents in the media (5). Typically, at small separations quantum (zero-point) fluctuations shape the force, whereas at separations large compared to the thermal wavelength $\lambda_T = \hbar c/(2\pi k_B T)$ (approximately 1.2 $\mu$m at room temperature $T = 300K$), thermal effects dominate and give rise to equilibrium or non-equilibrium Casimir forces (5, 6, 7).

Over the last two decade, there has been considerable progress in precision measurements of heat transfer and Casimir forces at sub-micron scale. [1] This growing interest can be attributed to the fact that heat transfer measurements are directly connected to scanning tunneling microscopy, and scanning thermal microscopy, under ultra-high vacuum conditions (8, 9), whereas Casimir forces must be accounted for in the fabrication of micro and nano electromechanical devices actuated by electrical bias (where this force is dominant). Thus, the development of theoretical tools that allow treatment of systems out of thermal equilibrium is of particular relevance. Unlike the equilibrium case, such powerful tools of statistical physics as entropy and Helmholtz free energy (containing information about forces and torques) cannot be applied to systems out of thermal equilibrium. One approach is to apply the formalism of fluctuational electrodynamics (FE) introduced by Rytov (10) in the 1950's (11, 12). As discussed in the Section 2, Rytov's formalism is based on the assumption of local equilibrium for each object, which is characterized by a temperature and dielectric response. The study of multiple objects of different shapes and material properties can be achieved by use of scattering methods (13, 14, 15, 16), as explored in Sections 3. Merging Rytov and scattering formalisms enables to compactly represent radiation and forces in terms of scattering operators of the individual objects (13, 14, 15, 16), as demonstrated in Section 4. Several applications of the results demonstrating the importance of size, shape, and geometry are provided in Section 5. Overall, the goal of this review is to illustrate the following novel twists on textbook results for heat and force, as summarized in Fig. 1:

**1. Heat radiation** by objects with sizes smaller or comparable to the thermal wavelength is not accurately described by the Stefan-Boltzmann law. This occurs due to interference effects between the object and the emitted radiation, such that its emissivity and absorptivity depend sensitively on its size and shape. Additionally, if the object is smaller than the penetration (skin) depth, the emitted power is proportional to the object's volume, rather than its surface area. For example, the top row of Fig. 1 depicts radiation from a silica nanofiber of diameter 500nm, together with the prediction of Eq. (41) below, displaying good agreement. Indeed, a naive computation using the Stefan-Boltzmann law (also shown in the figure) does not correctly predict the measured data. Theoretical studies of these effects have been carried out for spheres, plates and cylinders, [2] with the scattering formalism (10, 20, 21, 22, 23, 24, 25, 26). Furthermore, recent studies on superscattering properties of subwavelength nanostructures (e.g. nanorods) (27) make such systems potential candidates for efficient heat transfer applications.

**2. Heat transfer** between multiple objects is further complicated by the appearance of another scale, the separation between objects, yielding non-trivial effects if it is smaller or comparable to the thermal wavelength. Indeed, over 40 years ago van Hove and Polder used Fluctuational Electrodynamics to predict that radiative heat transfer between objects separated by a vacuum gap can exceed the blackbody limit (28), due to evanescent electromagnetic fields decaying exponentially into the vacuum. After the pioneering experiments in the late 60's and early 70's (29, 30), the enhancement of heat transfer in the near-field regime (generally denoting separations small compared to the thermal wavelength) has only recently been conclusively verified experimentally (18, 31, 32, 33, 34, 35, 36). The results of one such experiment are depicted in the middle row of Figure 1. Theoretically, heat transfer has been considered for a limited number of shapes: parallel plates (8, 28, 37, 38), a dipole or sphere in front a plate (23, 39, 40), two dipoles or spheres (39, 41, 42), and a cone in front of a plate (43). The scattering formalism has been successfully exploited (6, 10, 23, 44, 45, 46, 16) in this context. Although powerful numerical techniques (43, 47, 48) exist for arbitrary geometries, analytical computations are limited to planar, cylindrical, spherical and ellipsoidal cases (49, 50).

**3. Casimir forces** *in equilibrium* have been subject of considerable theoretical investigation, and were finally confronted with several high precision experiments in the 1990s (51, 52, 53). The interested reader is referred to the extensive literature on the subject, a sampling of which is in references (3, 54, 55, 56, 57). The bottom row of Fig. 1 depicts a

---

[1] Parts of this review are taken with permission from the doctoral thesis of Dr. V. Golyk (MIT, 2014).

---

[2] While the results are exact for plates and spheres, the radiation by a cylinder is obtained within certain approximations (20, 21, 22)

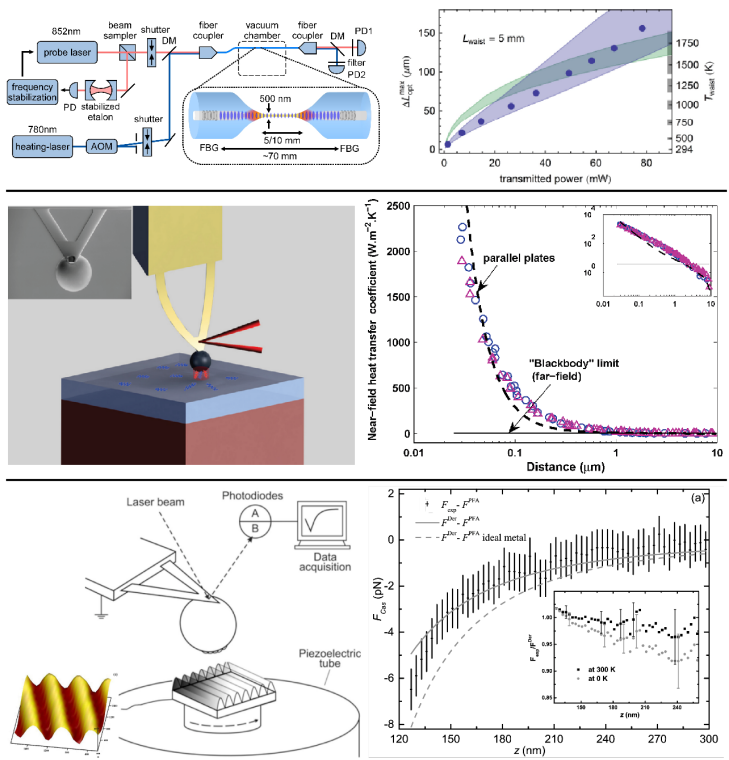

**Figure 1**

**Top row:** Thermal radiation from a silica nanofiber (diameter 500nm) from Ref. (17), measured through cooling rates of the fiber. The green bands are predictions from Planck's law, the purple bands from theory presented in Section 5, Eq. (41). The width of the bands arises through uncertainty of the fiber's dielectric properties. **Middle row:** Heat transfer from a heated glass bead (diameter of ∼100 microns) to a flat glass surface (from Ref. (18)) exceeds the Stefan-Boltzmann law (solid line) by orders of magnitude at separations below 0.1 micron. **Bottom row:** The equilibrium Casimir force between a corrugated surface and a corrugated sphere (19) is very well described by a corresponding theory (solid line).

recent example of Casimir force measurements between a corrugated surface and a corrugated sphere. (19) Even for this complicated geometry precise comparisons between theory and experiment are possible. There have been comparatively few studies of the Casimir force between objects at different temperatures. The force due to radiation pressure at large distances is modified by near field effects at separations less than the thermal wavelength. Non-equilibrium forces were theoretically predicted, initially between two plates or plate and an atom (12, 58, 59), and were later extended to other shapes with the scattering formalism (16, 44, 46). There has so far been only one experiment on the Casimir-Lifshitz force out-of-equilibrium (60), involving an ultra-cold atomic cloud at a distance of a few microns from a dielectric substrate. It is hoped that there will be further high precision experiments of non-equilibrium forces between solid objects, and indeed there has been a proposal (61, 62, 63, 64) to employ differential measurements (65, 66) to this end.

## 2. FLUCTUATIONAL ELECTRODYNAMICS

The goal of Fluctuational Electrodynamics is to study the phenomena originating from fluctuations of the electromagnetic field in macroscopic (globally neutral) bodies, importantly in the near-field regime of small distances. A first-principle analysis starting from the known quantum-mechanical interactions of the individual atoms comprising the two bodies is feasible only for rarefied objects. The microscopic approach becomes unpractical for macroscopic bodies, due to complicated inductive many-body effects that arise for particle densities characteristic of condensed matter. Indeed, for separations much larger than interatomic distances, a continuum description is appropriate, and interactions can be regarded as occurring though the fluctuating electromagnetic field which is always present in the interior of absorbing media, as a result of the endless quantum and thermal motions of its constituent charges. A key observation is that this field is not simply confined to the interior of the bodies, but stretches out well beyond, partly in the form of propagating waves and partly as exponentially decaying evanescent waves. Consequently, absorbing bodies are surrounded by an indestructible envelope of electromagnetic fluctuations. When two or more bodies are in close proximity, their electromagnetic envelopes overlap and mediate energy or momentum transfer between them, even in the absence of direct physical contact. For computing interactions, one expects that the relevant wavelengths of the fluctuating electromagnetic field should be comparable to their macroscopic scales and separation, and thus a description in terms of *macroscopic* electromagnetic features of the bodies should be feasible. The macroscopic approach outlined here has a large degree of generality: it naturally embodies thermal effects as well as retardation phenomena associated with the finite speed of light. As a check, the method should (and does) reproduce results obtained by considering quantum mechanical interactions of individual atoms, in the limit of dilute media.

To determine the statistical features of the electromagnetic field surrounding an absorbing body, Rytov (67, 10) postulated that the fluctuations are sourced by randomly fluctuating currents of density $\mathbf{j}(\mathbf{r}, t)$ existing in the interior of the body. Physically, the current $\mathbf{j}(\mathbf{r}, t)$ can be thought of as originating from the quantum and thermal random motions of the charges that constitute the body, i.e. conduction electrons in a conductor, bound electrons in a dielectric, or ions in a polar dielectric. The random field $\mathbf{j}(\mathbf{r}, t)$ is analogous to the random force that is introduced in the theory of Brownian motion. Consistent with the macroscopic character of the theory, it is assumed that the electromagnetic field generated by the random current obeys the macroscopic classical Maxwell's Equations. For a monochromatic field (with time dependence proportional to $e^{-i\omega t}$) in a dielectric, nonmagnetic inhomogeneous medium, these Equations read:

$$\nabla \times \mathbf{E}(\mathbf{r}, \omega) = i\frac{\omega}{c}\mathbf{H}(\mathbf{r}, \omega), \tag{1}$$

$$\nabla \times \mathbf{H}(\mathbf{r}, \omega) = -i\,\varepsilon(\mathbf{r}, \omega)\frac{\omega}{c}\mathbf{E}(\mathbf{r}, \omega) + \frac{4\pi}{c}\mathbf{j}(\mathbf{r}, \omega), \tag{2}$$

where $\varepsilon(\mathbf{r}, \omega)$ is the complex electric permittivity of the medium. The above Equations hold in the interior of each body, and must be supplemented with the usual boundary conditions on the fields $\mathbf{E}(\mathbf{r},\omega)$ and $\mathbf{H}(\mathbf{r},\omega)$ on the surfaces of the bodies. The presence of the random source $\mathbf{j}(\mathbf{r},\omega)$ gives to Equations 2 the character of a Langevin Equation for the electromagnetic field. The random current has zero mean, $\langle \mathbf{j}(\mathbf{r},\omega)\rangle = 0$, and its statistical properties are described by its two-point correlation function, providing the (quantum and thermal) average of the product of components of $\mathbf{j}(\mathbf{r},\omega)$ at two different points $\mathbf{r}$ and $\mathbf{r}'$ inside the body. By the very nature of the macroscopic approach, in which atomic distances are considered to be negligibly small, this correlation has the character of a delta-function $\delta(\mathbf{r}-\mathbf{r}')$. According to Rytov, the current correlator has the expression:

$$\langle j_i(\mathbf{r},\omega) j_k^*(\mathbf{r}',\omega')\rangle = \frac{\hbar\omega^2}{2}\coth\left(\frac{\hbar\omega}{2k_B T}\right)\mathrm{Im}\,\varepsilon(\mathbf{r},\omega)\delta_{ik}\delta(\mathbf{r}-\mathbf{r}')\delta(\omega-\omega'). \qquad (3)$$

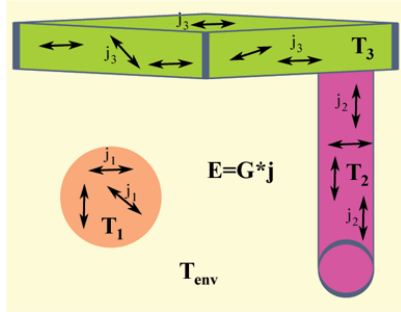

**Figure 2**
Fluctuations of the electromagnetic field outside several bodies, sourced by the fluctuating currents in each body.

Here, $i$ and $k$ label vector components, and $T$ is the temperature. To be precise, since in quantum theory the current density is an operator, the correlator is better understood as the average of the symmetrized product of the components of the current operator. Eq. 3 is a form of the fluctuation dissipation theorem (68), relating fluctuations of the currents to the dissipative response of the system, here represented by a scalar dielectric function $\varepsilon(\mathbf{r},\omega)$. For the benefit of the reader, we present in a separate Appendix an elementary derivation of Eq. 3, based on a semi-classical model of the dielectric as a collection of damped harmonic oscillators. The identity,

$$\frac{1}{2}\hbar\omega\coth\left(\frac{\hbar\omega}{2k_B T}\right) = \hbar\omega\left(\frac{1}{2} + \frac{1}{\exp(\hbar\omega/k_B T)-1}\right), \qquad (4)$$

recasts the correlator in Eq. 3 as the sum of a purely quantum contribution, represented by the first term between the brackets, and a thermal contribution which is proportional to the Bose-Einstein average occupation.

The electromagnetic field outside the body is obtained by solving Maxwell Equations 2, with $\mathbf{j}(\mathbf{r},\omega)$ as a source. This is best done using the Green's function formalism. The dyadic Green's function $\mathcal{G}_{ik}(\omega;\mathbf{r},\mathbf{r}')$ provides the electric field $E_i(\mathbf{r},\omega)$ at point $\mathbf{r}$, generated by a point-like current source located at the point $\mathbf{r}'$ inside the dielectric body. To preserve causality, $\mathcal{G}_{ik}(\omega;\mathbf{r},\mathbf{r}')$ is subjected to retarded boundary conditions. It is worth pointing out that the Green's function automatically takes into account non-additive many-body effects associated with the induced currents caused by the presence of the point-like source at $\mathbf{r}'$. Using the Green's function, the electric field generated by the continuous distribution of currents is expressed as:

$$E_i(\mathbf{r},\omega) = \frac{4\pi\mathrm{i}\,\omega}{c^2}\int d^3\mathbf{r}'\,\mathcal{G}_{ik}(\omega;\mathbf{r},\mathbf{r}')\,j_k(\mathbf{r}',\omega)\;. \tag{5}$$

Of course, since the random current has zero mean, the average electric field vanishes. However, the observables of interest are quadratic in the field components. For example, the force acting on a body is obtained by integrating the Maxwell tensor $T_{ij}$

$$T_{ij} = \frac{1}{4\pi}\left[E_i E_j + H_i H_j - \frac{1}{2}\delta_{ij}\left(E^2 + H^2\right)\right]\;, \tag{6}$$

over a closed surface drawn in the empty space outside the body, and enclosing the body in its interior, while the radiated power received by the body is obtained by integrating over such a surface the Poynting vector

$$\mathbf{S} = \frac{c}{4\pi}\mathbf{E}\times\mathbf{H}\;. \tag{7}$$

The average value of any observable that is quadratic in the electromagnetic field can be computed starting from the correlator of the electric field $\langle E_i(\mathbf{r},t)E_j(\mathbf{r}',t')\rangle$ (it follows from the first Maxwell Equation that correlators involving the magnetic field can be computed from the electric-field correlator by taking suitable curls). It follows from Equations 3 and 5 that the Fourier transform of $\langle E_i(\mathbf{r},t)E_j(\mathbf{r}',t')\rangle$ does not vanish and has the expression:

$$\langle E_i(\mathbf{r},\omega)E_j^*(\mathbf{r}',\omega')\rangle = \frac{8\pi\hbar\omega^4}{c^4}\coth\left(\frac{\hbar\omega}{2k_B T}\right)\delta(\omega-\omega') \tag{8}$$

$$\times \int d^3\mathbf{r}''\,\mathcal{G}_{ik}(\omega;\mathbf{r},\mathbf{r}'')\,\mathrm{Im}\,\varepsilon(\omega,\mathbf{r}'')\mathcal{G}_{kj}^*(\omega;\mathbf{r}'',\mathbf{r}'). \tag{9}$$

### 2.1. Systems in Global Thermal Equilibrium

When the system is in global thermal equilibrium at a temperature $T$, the integral over space in the r.h.s of the above Equation can be expressed in terms of the imaginary part of the Green's function, using the following important property of the Green's function:

$$\frac{\omega^2}{c^2}\int d^3\mathbf{r}''\,\mathcal{G}_{ik}(\omega;\mathbf{r},\mathbf{r}'')\,\mathrm{Im}\,\varepsilon(\omega,\mathbf{r}'')\mathcal{G}_{kj}^*(\omega;\mathbf{r}'',\mathbf{r}') = \mathrm{Im}\,\mathcal{G}_{ij}(\omega;\mathbf{r},\mathbf{r}')\;. \tag{10}$$

Inserting the above relation into the r.h.s. of Eq. 9, leads to

$$\langle E_i(\mathbf{r},\omega)E_j^*(\mathbf{r}',\omega')\rangle = \frac{8\pi\hbar\omega^2}{c^2}\coth\left(\frac{\hbar\omega}{2k_B T}\right)\delta(\omega-\omega')\,\mathrm{Im}\,\mathcal{G}_{ij}(\omega;\mathbf{r},\mathbf{r}'). \tag{11}$$

Equation 11 is a remarkable formula, indicating that in thermal equilibrium the correlator of the electric field at points *outside* the bodies is determined solely from knowledge of the Green's function in their exterior. Indeed, it has been shown (69, 70) that Eq. 11 can be directly obtained from the fluctuation-dissipation theorem by regarding the Green's function itself as a response of the system to external sources outside the bodies, without resort to hypothetical random currents in their interior. Another remarkable feature of the equilibrium is that the retarded Green's function of a passive medium is analytical in the upper complex frequency plane (71). By virtue of this analyticity, the equal-time correlator of the electric field $\langle E_i(\mathbf{r},t)E_j(\mathbf{r}',t)\rangle$ can be expressed, after a Wick-rotation of the frequency axis, as a sum over the imaginary poles $\omega_n = 2\pi i n k_B T/\hbar \equiv i\xi_n$, for $n = 0, 1, \ldots$ (the so-called Matsubara frequencies) of the hyperbolic cotangent in Eq. 11, as

$$\langle E_i(\mathbf{r},t)E_j(\mathbf{r}',t)\rangle = \frac{4k_B T}{c^2} {\sum_n}' \xi_n^2 \mathcal{G}_{ij}(i\xi_n;\mathbf{r},\mathbf{r}')\,, \qquad (12)$$

where the prime in the sum implies that the $n = 0$ term has to taken with weight $1/2$. This is a very useful formula, because Green's functions are much better behaved along the imaginary axis than along the real frequency axis, which makes Eq. 12 much easier to estimate numerically in practical applications.

Equation 11 shows how the spectral density of the fluctuating electromagnetic field is modified by the presence of one or more bodies, and in principle it allows to determine the fluctuation of the electric field for any number of dielectric bodies of any shape, in global thermal equilibrium. The difficulty resides of course in the computation of the Green's function $\mathcal{G}_{ij}(\omega;\mathbf{r},\mathbf{r}')$. For simple geometries and/or arrangements of the bodies, the Green's function can be worked out analytically, but in general its computation is very hard and can be only achieved numerically. An important general lesson about the influence of material properties on the equilibrium values of field correlators, and of the observables that can be computed based on them, can however be deduced from the imaginary-frequency formula in Eq. 12. Since the electric permittivity $\varepsilon(i\xi)$ of all causal materials is positive and monotonically decreasing along the imaginary frequency (71), equilibrium values of field correlators are substantially insensitive to the structure of the resonances that the dielectric function may display for real frequencies.

The simplest situation is that of an infinite homogeneous dielectric medium (71). The Green's function $\mathcal{G}_{ij}^{(0)}(\omega;\mathbf{r},\mathbf{r}')$ in this case is known analytically and the correlators of the electromagnetic field can be obtained in closed form. In the limit of a transparent medium, corresponding to $\mathrm{Im}\,\varepsilon \to 0$ (with $\mathrm{Re}\,\varepsilon > 0$), it is possible to verify that the spectral density of the electromagnetic energy density computed on the basis of Eq. 11 reproduces, after subtraction of the unphysical contribution of zero-point fluctuations, the spectral density of the black body radiation in a transparent medium. When finite geometries are considered, Eq. 11 can be used to study how the black body spectrum is modified by the shape and material properties of the cavity walls, influencing the spectral lines and lifetimes of atoms placed inside the cavity (69).

An important application of the equilibrium formulae in Equations 11 and 12 is to the Casimir effect (4). The Casimir effect is the tiny force between two neutral macroscopic polarizable bodies, that originates from quantum and thermal fluctuations of the electromagnetic field in the region of space bounded by the surfaces of the two bodies. As one of the rare manifestations of quantum mechanics at the macroscopic scale, like superconductivity or superfluidity, it has attracted considerable interest, also in view of its potential applications to nanotechnology, condensed matter physics, gravitation and cosmology (see, e.g. reviews in Refs. (3, 11, 55, 54, 57)). In his seminal work Casimir computed the force $F_C$ between two neutral perfectly conducting plane parallel plates of area $A$ at a distance $d$ in vacuum; by carefully summing the zero-point energies of the electromagnetic modes in the empty space between the plates, he found

$$F_C = \frac{\pi^2 \hbar c}{240} \frac{A}{d^4} \ . \tag{13}$$

The presence of Planck's constant clearly indicates the quantum origin of the Casimir force. Within the framework of fluctuational electrodynamics the Casimir force is estimated by integrating the Maxwell stress tensor in Eq. 6 across a closed surface $S$ drawn in vacuum and surrounding one of the two bodies. This in turn involves evaluating an integral on $S$ of field correlators evaluated at coinciding points $\mathbf{r} = \mathbf{r}'$ along $S$. Since the field correlators diverge at coincidence points, as the Green's function $\mathbf{G}(\omega; \mathbf{r}, \mathbf{r}')$ diverges in such limit, it is not immediately clear if a well defined force is to be obtained. That this is indeed so is recognized by considering the following decomposition of the Green's function at points outside the bodies:

$$\mathcal{G}_{ij}(\omega; \mathbf{r}, \mathbf{r}') = \mathcal{G}_{ij}^{(0)}(\omega; \mathbf{r}, \mathbf{r}') + \mathcal{G}_{ij}^{(\text{bodies})}(\omega; \mathbf{r}, \mathbf{r}'), \tag{14}$$

where as before $\mathcal{G}_{ij}^{(0)}(\omega; \mathbf{r}, \mathbf{r}')$ represents the free Green's function in infinite space, and $\mathcal{G}_{ij}^{(\text{bodies})}(\omega; \mathbf{r}, \mathbf{r}')$ is the contribution due to the presence of the bodies. In the coincidence limit $\mathbf{r} \to \mathbf{r}'$, only the free space Green's function diverges, while the body Green's function $\mathcal{G}_{ij}^{(\text{bodies})}(\omega; \mathbf{r}, \mathbf{r}')$ attains a finite limit (16). It is clear physically that the ill-defined contribution to the stress tensor originating from the free-space Green's function, being independent of the positions and material properties of the bodies, can be disregarded altogether, and that the Casimir force is determined solely by the well defined stresses associated with the body Green's function. A computation of the Casimir force along these lines was performed for the first time by Lifshitz (5) for the case of two plane-parallel semi-infinite dielectric slabs. In this simple geometry, the body Green's function $\mathcal{G}_{ij}^{(\text{bodies})}(\omega; \mathbf{r}, \mathbf{r}')$ can be easily determined analytically by working in a basis of plane waves propagating in the plane of the slab surfaces (16), and it can be expressed in terms of the Fresnel reflection coefficients of the slabs (72). The formula obtained by Lifshitz allowed for the first time to analyze the influence on Casimir pressure of real material properties of the bounding plates, like their finite reflectivity and temperature. The Lifshitz formula is routinely used today to interpret modern precision measurements of the Casimir force.

A major unresolved problem in Casimir physics concerns the magnitude of the thermal contribution to the Casimir force between two conducting plates. In essence the puzzle is about the role played by relaxation properties of conduction electrons in Lifshitz theory (55, 73). It turns out that Lifshitz formula predicts significantly different magnitudes for the thermal force depending on whether the optical data of the conductor are extrapolated towards zero frequency on the basis of the Drude model (which does take dissipation into account) or instead by the dissipationless plasma model of IR optics. In addition to predicting different magnitudes for the thermal force, it has been shown that the two prescriptions have important thermodynamic consequences: while the Drude prescription leads to a violation of Nernst heat theorem (in the idealized case of two conducting plates with a perfect crystal structure) (74), the plasma prescription violates the Bohr-van Leeuwen theorem of classical statistical physics (75, 76). It has been suggested in (77) that the reported violation of Nernst heat theorem by Drude metals may be due to a glassy state of quasi-static Foucault currents in Drude metals at low temperatures, which makes Nernst theorem not applicable. The experimental situation is contradictory. While several small distance experiments (78, 79, 80), probing separations below one micron, appear to be in agreement with the plasma model, and to rule out the larger thermal force predicted by the Drude model, a recent large-distance experiment in the wide range from 0.7 to 7.3 $\mu$m (81) has been interpretd as being in agreement with the Drude model. A novel experimental scheme based on differential Casimir force measurements (61, 62, 63) may lead to a definitive clarification of the problem. A very recent experiment (65, 66) based on this scheme, utilizing samples with alternating Au-Ni regions, appears to be very encouraging.

## 2.2. Systems out of Thermal Equilbrium

For out of equilibrium situations in which we can assign to each body (in its rest frame) a distinct temperature $T$, the basic assumption of fluctuational QED (10, 67) is that current fluctuations in each body still satisfy Eq. 3. Electric field fluctuations outside the bodies are obtained by summing contributions from the different objects, generalizing Eq. 9, and stress tensor and Poynting vector computed as in Equations 6 and 7. The technical difficulty is in expressing the Green's function for the collection of bodies, in terms of their individual Green's functions. A general remark is in order at this point. When studying problems of heat transfer, the contribution of vacuum quantum fluctuations (corresponding to the first term between the round brackets on the r.h.s. of Eq. 4) can be ignored altogether because, being independent of the bodies temperatures, vacuum fluctuations contribute nothing to energy transfers.

As a first step, it is instructive to consider the simple problem of a half-space filled with a homogeneous dielectric medium at temperature $T$. Working in a plane wave basis, the Green's function at points outside the dielectric is easily expressed in terms of its Fresnel reflection coefficients (16). One finds that the thermal fluctuating electromagnetic field outside the slab is composed of a superposition of propagating waves, and a near field contribution from evanescent waves that decay exponentially away from the slab surface. A straightforward computation of the average Poynting vector at large distances from the surface shows that Rytov's theory reproduces the well known Kirchhoff's law for the radiation of a surface (28). Unexpectedly, the thermal near field emitted by a surface has been shown to possess a remarkable degree of coherence (82), contrary to one's idea that thermal fluctuations are an incoherent phenomenon. Valuable insight is gained from the examination of the spectral density $u(\omega, z)$ of the energy density of the fluctuating electromagnetic field at a finite distance $z$ from the surface (83). It is found that for separations $z$ smaller than the characteristic thermal wavelength $\lambda_T = \hbar c/(2\pi k_B T)$ ($\lambda_T = 1.2$ $\mu$m at room temperature) the spectral density $u(\omega, z)$ is sharply different from its black body value, which is approached only for separations $z \gg \lambda_T$. In general, one finds that the energy density of the electromagnetic field increases monotonically as the surface is approached, achieving values that can be much larger than the black body limit. This enhancement originates from evanescent waves that exist in proximity of the surface, and can be spectacularly strong (several orders of magnitudes already for room temperature!) for materials supporting surface waves that can be thermally excited (like SiC). These simple observations lead one to expect that radiative heat transfer between two closely spaced surfaces at different temperatures may display novel features that cannot be seen for large separations.

The problem of radiative heat transfer for two plane-parallel homogeneous dielectric slabs at (arbitrary) different temperatures was first studied by Polder and Van Hove, using Rytov theory (28), extending a previous investigation by Rytov of the less realistic case of a heated dielectric slab facing an almost perfect mirror at zero temperature (10). The main finding of Polder and Van Hove is the strong increase of radiative heat transfer between two metallic surfaces for gap widths smaller than $\lambda_T$. Subsequent studies by other authors (83, 84) revealed that near field radiative heat transfer can in fact get enhanced by several orders of magnitude when the surfaces are covered by adsorbates or can support low-frequency surface plasmons, via a mechanism of resonant photon tunneling. The strong increase of the power of radiative transfer at the nanoscale opens up the possibility of important technological applications including heat-assisted magnetic recording (85), near-field thermophotovoltaics (86) and lithography (8).

Rytov theory out of thermal equilibrium has been recently used to investigate the Casimir force between two dielectric slabs at different temperatures in vacuum (12). Out of thermal equilibrium the Casimir-Lifshitz force displays remarkable features that disappear when the system is brought in a state of thermal equilibrium. These features originate from a peculiar contribution, $\bar{F}^{(\text{neq})}(T_1, T_2)$, to the non-equilibrium force, which is *antisymmetric* under an exchange of temperatures $T_1$ and $T_2$. Being antisymmetric in the body temperatures, this term can have either sign and can in principle be harnessed to tune the force both in strength and sign (87), and to realize self-propelling systems (6). Some of these remarkable phenomena shall be discussed in greater detail in Section 5.

## 3. SCATTERING THEORY

We have seen above that fluctuation induced effects can be described within Rytov's theory as emerging from fluctuating, random currents inside the bodies. Within this approach, it is sufficient to know the universal correlation function of those currents since they are related by the Green's function to the corresponding fields from which the relevant observables such as forces and energy fluxes can be obtained by means of the stress tensor. Hence, it is desirable to express the energy of a system of bodies that each carry a prescribed charge and current distribution in terms of these quantities. With this functional, one can then average over random currents to obtain the free energy or field correlation functions.

This observation leads us to the question of what determines the energy of a given set of bodies that carry charges and currents. We shall see below that the answer to this is given by scattering theory. Before we describe the general scattering approach, it is instructive to get familiar with the concept by looking at the energy of two distant conductors of arbitrary shape. Putting a charge on one conductor changes the potential of the other by an amount that depends on the geometrical configuration and the individual capacitances of the conductors. The potential $m_{12}$ to which conductor 2 is raised when a unit charge is placed on conductor 1 is known as *elastance*. Due to Green's reciprocity theorem, one has $m_{12} = m_{21}$ so that the elastance matrix $M = [m_{\alpha\beta}]$ is symmetric. Since the potentials of the conductors are given by $[V_1, V_2] = M[Q_1, Q_2]$ in terms of their charges, the electrostatic energy is given by the quadratic form $E = \frac{1}{2}[Q_1, Q_2]^t M[Q_1, Q_2]$. We shall see below that a similar quadratic form is obtained for the energy of a system of arbitrarily shaped dielectric bodies with (frequency dependent) fluctuating currents.

The elastance matrix contains information about the shapes and relative positions of the objects. Let the two conductors have capacitances $C_1$ and $C_2$ in isolation. When 1 is uncharged, and 2 with charge $Q_2$ is placed at a distance $d$ (which is assumed to be much larger than the linear dimension $R$ of the conductors) then the potential of 1 is raised to $Q_2/(4\pi d)$. Hence $m_{21} = 1/(4\pi d)$. On the nearer half of 1 a charge of opposite sign to $Q_2$ and magnitude $\sim C_1 Q_2/(4\pi d) \sim Q_2 R/d$ is induced (and an equal charge of equal sign is induced on the remote half). This leads at 2 to a dipole potential of the order $Q_2 R^2/d^3$ which can be ignored to leading order in $R/d$. Hence the potential $V_2$ is not affected by 1, and we obtain $m_{22} = V_2/Q_2 = C_2^{-1}$, and similarly for $m_{11}$, such that the elastance matrix is

$$M = \begin{pmatrix} C_1^{-1} & (4\pi d)^{-1} \\ (4\pi d)^{-1} & C_2^{-1} \end{pmatrix}. \tag{15}$$

We note that the self- and mutual capacitances are given by the inverse of the matrix $M$. This matrix shows two interesting features. The diagonal elements contain solely information about the shape of the objects (via their capacitances), while the off-diagonal elements depend only their relative position.

Now consider the general situation of arbitrarily shaped dielectric bodies that carry fluctuating currents $\mathbf{J}_j(\omega)$ where $j$ numbers the bodies and $\omega$ is the frequency. The equivalent of the capacitance is the scattering amplitude, also known as the $T$-operator $\mathbb{T}$ (88). It measures the amount of fluctuating currents that are induced on a body in response to an incident electromagnetic field. We briefly review the key results from scattering theory. The general solution of the wave equation

$$(\mathbb{H}_0 - \mathbb{V}(\omega, \mathbf{x})) \mathbf{E}(\omega, \mathbf{x}) = \frac{\omega^2}{c^2} \mathbf{E}(\omega, \mathbf{x}), \tag{16}$$

with differential and potential operators

$$\mathbb{H}_0 = \boldsymbol{\nabla} \times \boldsymbol{\nabla} \times, \tag{17}$$

$$\mathbb{V}(\omega, \mathbf{x}) = -\mathbb{I} \frac{\omega^2}{c^2} \left(1 - \epsilon(\omega, \mathbf{x})\right) - \boldsymbol{\nabla} \times \left(\frac{1}{\mu(\omega, \mathbf{x})} - 1\right) \boldsymbol{\nabla} \times, \tag{18}$$

is given by the Lippmann-Schwinger equation

$$\mathbf{E} = \mathbf{E}_0 + \mathbb{G}_0 \mathbb{V} \mathbf{E}. \tag{19}$$

Here $\mathbb{G}_0$ is the free electromagnetic tensor Green's function, and the homogeneous solution $\mathbf{E}_0$ obeys the wave equation with $\mathbb{V} = 0$. One can iteratively substitute for $\mathbf{E}$ in Eq. (19) to obtain the formal expansion

$$\begin{aligned} \mathbf{E} &= \mathbf{E}_0 + \mathbb{G}_0 \mathbb{V} \mathbf{E}_0 + \mathbb{G}_0 \mathbb{V} \mathbb{G}_0 \mathbb{V} \mathbf{E} - \ldots \\ &\equiv \mathbf{E}_0 + \mathbb{G}_0 \mathbb{T} \mathbf{E}_0, \end{aligned} \tag{20}$$

where the electromagnetic $\mathbb{T}$-operator is defined as

$$\mathbb{T} = \mathbb{V} \frac{\mathbb{I}}{\mathbb{I} - \mathbb{G}_0 \mathbb{V}} = \mathbb{V} \mathbb{G} \mathbb{G}_0^{-1}, \tag{21}$$

and $\mathbb{G}$ is the Green's function of the wave equation in Eq. (16), given explicitly by

$$\mathbb{G} = \mathbb{G}_0 + \mathbb{G}_0 \mathbb{T} \mathbb{G}_0. \tag{22}$$

We note that $\mathbb{T}$, $\mathbb{G}_0$, and $\mathbb{G}$ are all functions of frequency $\omega$ and non-local in space. As can be seen from expanding $\mathbb{T}$ in Eq. (21) in a power series, $\mathbb{T}(\omega, \mathbf{x}, \mathbf{x}') = \langle \mathbf{x} | \mathbb{T}(\omega) | \mathbf{x}' \rangle$ is zero whenever $\mathbf{x}$ or $\mathbf{x}'$ are not located on an object, i.e., where $\mathbb{V}(\omega, \mathbf{x})$ is zero.

The matrix elements of the $\mathbb{T}$-operator are given by the scattering amplitude $\mathcal{T}$ in a given basis, e.g., vector spherical waves. We consider a scattering process in which a regular wave $\mathbf{E}_\alpha^{\text{reg}}$ interacts with a body and scatters outward in form of an outgoing wave $\mathbf{E}_\beta^{\text{out}}$. Here $\alpha$ and $\beta$ label basis elements. We choose a convenient "scattering origin" *inside* r the body, consistent with any symmetries of the problem if possible.

To find the field $\mathbf{E}$ at a coordinate $\mathbf{x}$ far enough *outside* the body, we employ Eq. (20) in position space and the usual expansion of $\mathbb{G}_0$ in terms of regular and outgoing waves,

$$\mathbf{E}(\omega, \mathbf{x}) = \mathbf{E}_\alpha^{\text{reg}}(\omega, \mathbf{x}) + \sum_\beta \mathbf{E}_\beta^{\text{out}}(\omega, \mathbf{x}) \mathcal{T}_{\beta\alpha}(\omega). \tag{23}$$

where the scattering amplitude is defined by

$$\mathcal{T}_{\beta\alpha}(\omega) = i \iint \mathbf{E}_\beta^{\text{reg}*}(\omega, \mathbf{x}') \cdot \mathbb{T}(\omega, \mathbf{x}', \mathbf{x}'') \mathbf{E}_\alpha^{\text{reg}}(\omega, \mathbf{x}'') d\mathbf{x}' d\mathbf{x}''. \tag{24}$$

In analogy to the electrostatic case in Eq. (15), we can now express the kernel of the action for the fluctuating currents at fixed frequency on an arbitrary number of bodies,

$$S(\omega) = \frac{1}{2} \mathbf{J}_j \mathbb{M}_{jk} \mathbf{J}_k, \tag{25}$$

in terms of the inverse of the T-operators $\mathbb{T}_j$ of the individual bodies, replacing the inverse capacitances on the diagonal,

$$\mathbb{M} = \begin{pmatrix} -\mathbb{T}_1^{-1} & \mathbb{U}^{12} & \mathbb{U}^{13} & \cdots \\ \mathbb{U}^{21} & -\mathbb{T}_2^{-1} & \mathbb{U}^{23} & \cdots \\ \cdots & \cdots & \cdots & \cdots \end{pmatrix}. \tag{26}$$

We have not specified yet the off-diagonal operators of this kernel. Again, in analogy to the electrostatic case, they must describe the interaction of the currents on different objects. Hence, the operators $\mathbb{U}^{\alpha\beta}$ must be related to the free Green's function $\mathbb{G}_0$. The latter can be expanded as

$$\mathbb{G}_0(\omega, \mathbf{x}, \mathbf{x}') = i \sum_\alpha \begin{cases} \mathbf{E}_\alpha^{\text{out}}(\omega, \xi_1, \xi_2, \xi_3) \otimes \mathbf{E}_\alpha^{\text{reg}*}(\omega, \xi_1', \xi_2', \xi_3') & \text{if } \xi_1(\mathbf{x}) > \xi_1'(\mathbf{x}') \\ \mathbf{E}_\alpha^{\text{reg}*}(\omega, \xi_1, \xi_2, \xi_3) \otimes \mathbf{E}_\alpha^{\text{out}}(\omega, \xi_1', \xi_2', \xi_3') & \text{if } \xi_1(\mathbf{x}) < \xi_1'(\mathbf{x}') \end{cases}, \tag{27}$$

where one of the spatial coordinates is identified as the "radial" variable and treated differently from the remaining coordinates. We let $\xi_1$ represent this coordinate and denote the remaining coordinates as $\xi_2$ and $\xi_3$. We introduce the "outgoing" solution in $\xi_1$, which is in the same scattering channel as the corresponding regular solution but obeys outgoing wave boundary conditions as $\xi_1 \to \infty$. It is linearly independent of the regular solution. Since the $\mathbb{T}$-operators for the individual bodies in Eq. (26) are defined relative to the basis appropriate for each body, one has to translate the scattering solution of one body to the basis of the other body. We assume that all objects are placed outside of each other, see Fig. 3(b). The outgoing solutions form a complete set independent of the origin used to define the decomposition into a basis of partial waves. Let $\{\mathbf{E}_\beta^{\text{reg}}(\omega, \mathbf{x}_j)\}$ be the regular solutions expressed with respect to the origin of coordinates appropriate to object $j$, $\mathcal{O}_j$. Except in a region that contains the origin $\mathcal{O}_i$, where $\mathbf{E}_\alpha^{\text{out}}(\omega, \mathbf{x}_i)$ is singular, one can expand an outgoing solution in terms of the $\{\mathbf{E}_\beta^{\text{reg}}(\omega, \mathbf{x}_j)\}$,

$$\mathbf{E}_\alpha^{\text{out}}(\omega, \mathbf{x}_i) = \sum_\beta \mathbb{U}_{\beta\alpha}^{ji}(\omega, \mathbf{X}_{ji}) \mathbf{E}_\beta^{\text{reg}}(\omega, \mathbf{x}_j), \text{ for } \mathbf{x} \notin N(\mathcal{O}_i), \tag{28}$$

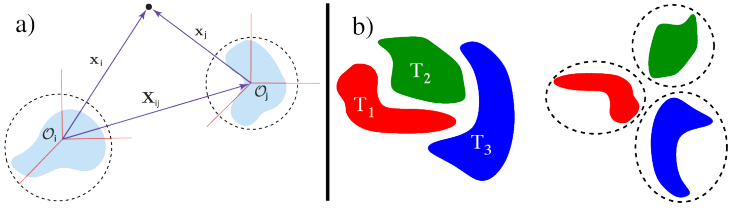

**Figure 3**

(a) Geometry of the configuration of two objects. The dotted lines show surfaces separating the objects on which the radial variable is constant. The translation vector $\mathbf{X}_{ij} = \mathbf{x}_i - \mathbf{x}_j = -\mathbf{X}_{ji}$ describes the relative positions of the two origins. (b) The operator formalism is completely general, and applies e.g. to the configuration on the left hand side. The right hand side shows a configuration that allows a representation in partial waves, because the enclosing spheres or ellipsoids do not overlap.

where $\mathbf{X}_{ij} = -\mathbf{X}_{ji} = \mathbf{x}_i - \mathbf{x}_j$ is shown in Fig. 3(a). Note that $\mathbf{x}_i$ and $\mathbf{x}_j$ refer to the same position in space, $\mathbf{x}$, expressed as the displacement from different origins. Here $N(\mathcal{O}_i)$ is a neighborhood of the origin $\mathcal{O}_i$. This expansion defines the matrix elements of the operator $\mathbb{U}_{\beta\alpha}$ is a given basis. We note that while the operator formalism in terms of $\mathbb{G}_0$ and $\mathbb{T}$ is fully general, and expansion in partial waves usually imposes constraints on the relative position of the bodies, see Fig. 3(b). There must exist an origin and a separable coordinate system so that for all points $\mathbf{x}$ in one object and $\mathbf{x}'$ in another object, $\xi_1(\mathbf{x})$ is always greater than $\xi_1(\mathbf{x}')$, or vice versa. Having $\xi_1(\mathbf{x}) > \xi_1(\mathbf{x}')$ ensures that the Green's function is always evaluated by letting $\mathbf{x}$ be the argument of the outgoing wave function and $\mathbf{x}'$ be the argument of the regular wave function. Therefore, any two objects must be separated by a surface defined by the locations $\mathbf{x}$ where $\xi_1(\mathbf{x})$ is constant.

This completes the construction of the kernel $\mathbb{M}$ which can be viewed in a given basis as the interaction matrix between multipole moments on the bodies. For a similar, earlier approach based on a multipole expansion for specific shapes like spheres and cylinders see Ref. (89). With this energy functional for the fluctuating currents on a set of arbitrarily shaped dielectric objects, the free energy and field correlation functions of the system can be computed by averaging over the currents with the corresponding Boltzmann weight. The first application of this approach has been the computation of equilibrium Casimir interaction (13, 15). The Casimir energy can be written as

$$\mathcal{E}_0 = -\frac{\hbar c}{2\pi} \int_0^\infty d\kappa \, \log Z(\kappa), \tag{29}$$

where we have Wick rotated from real to the imaginary frequencies $\kappa$, and $Z(\kappa)$ is the partition function that corresponds to the action of Eq. (25) for the currents. After integrating over the currents, we obtain

$$\mathcal{E} = \frac{\hbar c}{2\pi} \int_0^\infty d\kappa \, \log \det(\mathbb{M}\mathbb{M}_\infty^{-1}), \tag{30}$$

where $\mathbb{M}_\infty^{-1}$ is a block diagonal matrix diag($\mathbb{T}_1 \ \mathbb{T}_2 \ \cdots$). For the case of two objects, the expressions simplifies to

$$\mathcal{E} = \frac{\hbar c}{2\pi} \int_0^\infty d\kappa \, \log \det \left(\mathbb{I} - \mathbb{T}_1 \mathbb{U}^{12} \mathbb{T}_2 \mathbb{U}^{21}\right). \tag{31}$$

At nonzero temperature $T$ the integral $\frac{\hbar c}{2\pi} \int_0^\infty d\kappa$ is replaced by $k_B T \sum_n'$, where $\kappa_n = \frac{2\pi n k_B T}{\hbar c}$ with $n = 0, 1, 2, 3 \ldots$ is the $n$th Matsubara frequency. A careful analysis of the derivation shows that the zero frequency mode is weighted by $1/2$ compared to the rest of the terms in the sum; this modification of the sum is denoted by a prime on the summation symbol.

The scattering approach presented here builds on a range of previous related work, an inevitably incomplete subset of which is briefly reviewed here: Scattering theory methods were first applied to the parallel plate geometry, when Kats reformulated Lifshitz theory in terms of reflection coefficients (72). Jaekel and Reynaud derived the Lifshitz formula using reflection coefficients for lossless infinite plates (90) and Genet, Lambrecht, and Reynaud extended this analysis to the lossy case (91). Lambrecht, Maia Neto, and Reynaud generalized these results to include non-specular reflection (92).

Around the same time as Kats's work, Balian and Duplantier developed a multiple scattering approach to the Casimir energy for perfect metal objects and used it to compute the Casimir energy at asymptotically large separations (93, 94) at both zero and nonzero temperature. In their approach, information about the conductors is encoded in a local surface scattering kernel, whose relation to more conventional scattering formalisms is not transparent, and their approach was not pursued further at the time. One can find multiple scattering formulas in an even earlier article by Renne (95), but scattering is not explicitly mentioned, and the technique is only used to rederive older results.

Another scattering-based approach has been to express the Casimir energy as an integral over the density of states of the fluctuating field, using the Krein formula (96, 97, 98) to relate the density of states to the $\mathcal{S}$-matrix for scattering from the ensemble of objects. This $\mathcal{S}$-matrix is difficult to compute in general. In studying many-body scattering, Henseler and Wirzba connected the $\mathcal{S}$-matrix of a collection of spheres (99) or disks (100) to the objects' individual $\mathcal{S}$-matrices, which are easy to find. Bulgac, Magierski, and Wirzba combined this result with the Krein formula to investigate the scalar and fermionic Casimir effect for disks and spheres (101, 102, 103). Casimir energies of solitons in renormalizable quantum field theories have been computed using scattering theory techniques that combine analytic and numerical methods (104).

Bordag, Robaschik, Scharnhorst, and Wieczorek (105, 106) introduced path integral methods to the study of Casimir effects and used them to investigate the electromagnetic Casimir effect for two parallel perfect metal plates. Li and Kardar used similar methods to study the scalar thermal Casimir effect for Dirichlet, Neumann, and mixed boundary conditions (107, 108). The quantum extension was subsequently applied to the quantum electromagnetic Casimir effect by Emig, Hanke, Golestanian, and Kardar, who studied the Casimir interaction between plates with roughness (109) and between deformed plates (110). (Techniques developed to study the scalar Casimir effect can be applied to the electromagnetic case for perfect metals with translation symmetry in one spatial direction, since then the electromagnetic problem decomposes into two scalar ones.) Finally, the path integral approach was connected to scattering theory by Emig and Buescher (111).

Closely related to the work we present here is that of Kenneth and Klich, who expressed the data required to characterize Casimir fluctuations in terms of the transition $\mathbb{T}$-operator for scattering of the fluctuating field from the objects (112). Their abstract representation made it possible to prove general properties of the sign of the Casimir force. In Refs. (113, 114), we developed a framework in which this abstract result can be applied to concrete calculations. In this approach, the $\mathbb{T}$-operator is related to the scattering amplitude for each object individually, which in turn is expressed in an appropriate basis of multipoles. In this approach, the objects can have any shape or material properties, as long as the scattering amplitude can be computed in a multipole expansion (or measured). The approach can be regarded as a concrete implementation of the proposal emphasized by Schwinger (115) that the fluctuations of the electromagnetic field can be traced back to charge and current fluctuations on the objects. This formalism has been applied and extended in a number of Casimir calculations (116, 117, 118, 119, 120, 121, 122).

Even though the scattering formalism outlined above allows in principle to compute the Casimir energy for arbitrary geometries of the surfaces, its practical implementation is difficult unless one considers surfaces of simple shapes (like spheres and planes) for which the scattering amplitude is known. For more general shapes, one has to resort either to fully numerical schemes (43, 47, 48) or to make use of approximations. For the experimentally relevant case of two gently curved surfaces at small separations, a simple approximation routinely used to interpret current Casimir experiments is constituted by the so-called proximity force approximation (PFA) (123), which provides the leading contribution of the Casimir energy. By using a recently proposed derivative expansion it has now become possible to estimate also the leading curvature corrections to the PFA formula (124, 125, 126). The theoretical solid line displayed in the bottom row of Fig. 1 was indeed computed using the derivative expansion.

## 4. MERGING RYTOV & SCATTERING THEORY

Having set out the elements of scattering theory, in this section we describe how this methodology can be merged with the Rytov formalism to address heat radiation, heat transfer, and (Casimir) interactions for a set of arbitrary bodies in thermal non-equilibrium. We assume that $N$ dielectric objects, labelled by $\alpha = 1, \ldots, N$, kept at time-independent, homogeneous temperatures $\{T_\alpha\}$, are placed into an environment (vacuum) at temperature $T_{\text{env}}$. The bodies are characterized by their electric and magnetic response $\epsilon(\omega; \mathbf{r}, \mathbf{r}')$ and $\mu(\omega; \mathbf{r}, \mathbf{r}')$, which can in general be nonlocal complex tensors, $\epsilon(\omega; \mathbf{r}, \mathbf{r}') = \varepsilon_{ij}(\omega; \mathbf{r}, \mathbf{r}')$, depending on frequency $\omega$. In this non-equilibrium stationary state, each object is assumed to be at local equilibrium, such that the current fluctuations within the object satisfy the fluctuation dissipation theorem. The general results for heat radiation, transfer, and force are captured by a set of formulas reproduced in the box at the end of this section. In the following we provide a compact derivation of how these results are obtained. The reader is referred to Ref. (16) for a detailed exposition.

In equilibrium situations, the interactions (Casimir forces and torques) between the bodies can be obtained from the dependence on position and orientation of the free energy of the system. For non-equilibrium phenomena like heat radiation and transfer, the free energy is not well-defined, and the correlation function $\mathbb{C}(\mathbf{r}, \mathbf{r}') = [\langle E_i(\mathbf{r}) E_j^*(\mathbf{r}') \rangle_\omega]_{i,j=1,2,3}$ of the total field is needed to compute the current-averaged components of the Maxwell tensor. We shall see below that even out of equilibrium the correlation function is a superposition of equilibrium correlation functions, each at the temperature of the body that generates the corresponding contribution to the total field.

In equilibrium, the field correlations can be expressed as the superposition

$$\mathbb{C}^{eq}(T) = a_0 \operatorname{Im} \mathbb{G} + \sum_\alpha \mathbb{C}^{sc}_\alpha(T) + \mathbb{C}^{\text{env}}(T), \tag{32}$$

with $a_0 \equiv \text{sgn}(\omega)\frac{4\pi\hbar\omega^2}{2c^2}$. Here the first term represents the zero-point fluctuations, the sum in the second term the radiation coming from the fluctuating sources inside the bodies, and the last term is the contribution of the environment,

$$\mathbb{C}^{\text{env}}(T) = -a(T)\mathbb{G}\,\text{Im}\left[\mathbb{G}_0^{-1}\right]\mathbb{G}^*, \tag{33}$$

with

$$a(T) \equiv \text{sgn}(\omega)\frac{8\pi\hbar\omega^2}{c^2}[\exp(\hbar|\omega|/k_B T) - 1]^{-1}, \tag{34}$$

containing the occupation number of modes with frequency $\omega$, the speed of light $c$, and Planck's constant $\hbar$. In general, $\mathbb{C}^{sc}_\alpha(T)$ depends on the shape and material of *all objects* since the field radiated by body $\alpha$ is scattered by all other bodies. For the moment, we consider a single body at temperature $T$ in a cold environment and neglect zero-point fluctuations. Using the scattering formalism, the correlations of the radiated field can be easily expressed just in terms of the free Green's function and the body's T-matrix $\mathbb{T}$, as

$$\mathbb{C}^{sc,iso}_\alpha(T) = a(T)\mathbb{G}_0\left[\frac{i}{2}(\mathbb{T}^* - \mathbb{T}) - \mathbb{T}\,\text{Im}[\mathbb{G}_0]\mathbb{T}^*\right]\mathbb{G}_0^*. \tag{35}$$

This result is modified by the presence of the other objects. Solving the Lippmann-Schwinger equation yields the total field $\mathbf{E}_\alpha$ in terms of the field $\mathbf{E}_{\alpha,iso}$ of object $\alpha$ alone as $\mathbf{E}_\alpha = \mathbb{O}_\alpha \mathbf{E}_{\alpha,iso}$, with the multiple scattering operator

$$\mathbb{O}_\alpha = (1 + \mathbb{G}_0\mathbb{T}_{\bar\alpha})\frac{1}{1 - \mathbb{G}_0\mathbb{T}_\alpha\mathbb{G}_0\mathbb{T}_{\bar\alpha}}, \tag{36}$$

that depends also on the T-operator $\mathbb{T}_{\bar\alpha}$ of all other bodies. The final result for the multi-body correlation function $\mathbb{C}^{sc}_\alpha(T)$ in Eq. (32) is then given by

$$\mathbb{C}^{sc}_\alpha(T_\alpha) = \mathbb{O}_\alpha\,\mathbb{C}^{sc,iso}_\alpha(T_\alpha)\,\mathbb{O}^\dagger_\alpha. \tag{37}$$

We can thus express the full correlation function in terms of the free Green's function and the T-operators. In non-equilibrium, we can eliminate the environment contribution and obtain the correlation function as a superposition of the radiation from the bodies at different temperatures $T_\alpha$ as

$$\mathbb{C}^{\text{neq}}(\{T_\alpha\}, T_{\text{env}}) = \mathbb{C}^{\text{eq}}(T_{\text{env}}) + \sum_\alpha \left[\mathbb{C}^{\text{sc}}_\alpha(T_\alpha) - \mathbb{C}^{\text{sc}}_\alpha(T_{\text{env}})\right]. \tag{38}$$

With the functions expressed in terms of the T-operators of the bodies, we can compute radiated heat rates and forces in terms of these operators. The heat $H$ emitted by a body is obtained by integrating the normal component of the Poynting vector $\mathbf{S}$ over a surface $\Sigma$ enclosing the body,

$$H = \oint_\Sigma \mathbf{S} \cdot \mathbf{n}, \tag{39}$$

with $\mathbf{S}(\mathbf{r}) = \frac{c}{4\pi}\int_{-\infty}^{\infty}\frac{d\omega}{2\pi}\langle\mathbf{E}(\mathbf{r}) \times \mathbf{B}^*(\mathbf{r})\rangle_\omega$.

Note that the forces on two bodies at different temperatures are not equal and opposite, and have to be obtained separately. The force on object $\alpha$ is found from the surface normal component of the stress tensor $\sigma$, integrated over a surface enclosing object $\alpha$, as

$$\mathbf{F}^{(\alpha)} = \text{Re}\oint_{\Sigma_\alpha} \sigma \cdot \mathbf{n}, \tag{40}$$

with $\sigma_{ij} = \int_{-\infty}^{\infty} \frac{d\omega}{8\pi^2} \langle E_i E_j^* + B_i B_j^* - \frac{1}{2} (|E|^2 + |B|^2) \delta_{ij} \rangle_\omega$.

So far, we used a basis-independent operator representation. These results hold for any geometry. For practical computations, the operators have to represented in a basis of partial waves. The traces of operators turn then into sums over matrix elements with respect to the partial wave indices. In the box below we present the general trace formulas for heat radiation, heat transfer, and forces both in the operator formalism and a basis of spherical vector waves (16). Note that in the latter case, the spheres enclosing the bodies cannot overlap, see also Fig. 3(b).

## 5. SIZE, SHAPE, AND GEOMETRY

### 5.1. Radiation

The thermal emission $H$ of a macroscopic, isolated object at temperature $T$ is well understood, going back to the theory of black body radiation. It is proportional to the surface area $A$ of the object, and reads (127, 2)

$$\frac{H}{A} = \sigma T^4 \epsilon(T), \tag{42}$$

with $\sigma = \pi^2 k_B^4/(60\hbar^3 c^2)$ the Stefan-Boltzmann constant. The emissivity $\epsilon$ is in general a function of temperature. The law in Eq. (42) accurately describes the energy emitted if the object (and its radii of curvature) is large compared to the emitted wavelengths, which in turn are centered around the thermal wavelength $\lambda_T$. In other words when, from the perspective of the wave, the surface is nearly planar. [Note that we are here omitting the issue of view factors, which gives a correction to Eq. (42) due to its shape, when part of the object's radiation is reabsorbed by itself, if surface parts face each other (like in a ring).] Apart from concavity, Eq. (42) turns inaccurate if the size of the object is comparable or small compared to $\lambda_T$. The radiation can then strongly depend on the materials in combination with shape, as we aim to demonstrate with two standard examples: A sphere and a cylinder (wire) of infinite length.

We start with a sphere made of glass (specifically, $SiO_2$), which is a well absorbing (and emitting) material, with otherwise mild optical properties. Its radiation is depicted by the upper red curve in Fig. 4, as a function of the radius of the sphere (23, 16). For very large $R$, the curve indeed approaches the value given by Eq. (42), where the emissivity of the glass type chosen is $\epsilon = 0.735$. The important length scale is indeed the thermal wavelength $\lambda_T$, so that approach to Eq. (42) is observed for $R \gg \lambda_T$. In the other extreme, for very small $R$, the emitted energy is not proportional to the surface area $A$, but (note that the $y$-axis is normalized by $A$) proportional to the *volume* of the sphere. This can be understood from the fact that in this limit, the sphere is almost transparent, so that the radiation emitted by any volume element of the sphere contributes. This is in contrast to a large sphere, where only volume elements close to the surface contribute, as the emission from elements deeper inside is reabsorbed before reaching the surface.

Due to the plain optical properties of glass, the curve is rather featureless, crossing over between the above two limits. More complex behavior is observed for well conducting materials such as gold (16). The lower red curve in Fig. 4 follows a similar trend for large $R$, approaching Eq. (42) in the same manner as glass (only here, the value of the emissivity is much smaller, $\epsilon = 0.0075$). For small $R$, the form is however very different compared to glass, and emission proportional to volume is not observed for any relevant values of $R$.

**Macroscopic bodies:** Radiate proportional to their surface area

**Black Body:** A body with emissivity of unity, marking the maximal value for planar surfaces.

**Thermal wavelength** $\lambda_T$: Dominant wavelength in the Planck radiation spectrum.

**Skin depth** $\delta$: Length scale of damping of waves inside matter. For glass ($SiO_2$), around 1 micron. For gold, a few nanometers.

> **TRACE FORMULAS**
>
> Below we summarize trace formulas for the various non-equilibrium quantities (16). The expressions are given both in the general operator notation, and in spherical wave representation in which the $\mathbb{T}$ operator and $\mathbb{G}_0$ become (infinitely dimensional) matrices $\mathcal{T}$ and $\mathcal{U}$.
>
> **Heat radiation of one object**
>
> The heat emitted by a body with T-matrix $\mathbb{T}$ at temperature $T$ is given by (with $\mathcal{S} = \mathcal{I} + 2\mathcal{T}$)
>
> $$H = \frac{2\hbar}{\pi}\int_0^\infty d\omega \frac{\omega}{e^{\frac{\hbar\omega}{k_B T}}-1}\text{Tr}\left\{\text{Im}[\mathbb{G}_0]\text{Im}[\mathbb{T}] - \text{Im}[\mathbb{G}_0]\mathbb{T}\text{Im}[\mathbb{G}_0]\mathbb{T}^*\right\} = \frac{\hbar}{2\pi}\int d\omega \frac{\omega}{e^{\frac{\hbar\omega}{k_B T}}-1}\text{Tr}\left[\mathcal{I} - \mathcal{S}\mathcal{S}^\dagger\right] \quad (41)$$
>
> **Heat transfer between two objects**
>
> The radiation emitted by body 1 at temperature $T_1$ and absorbed by body 2 is given by the transfer rate
>
> $$H_1^{(2)} = \frac{2\hbar}{\pi}\int_0^\infty d\omega \frac{\omega}{e^{\frac{\hbar\omega}{k_B T_1}}-1}\text{Tr}\left\{[\text{Im}[\mathbb{T}_2] - \mathbb{T}_2^*\text{Im}[\mathbb{G}_0]\mathbb{T}_2]\frac{1}{1-\mathbb{G}_0\mathbb{T}_1\mathbb{G}_0\mathbb{T}_2}\mathbb{G}_0[\text{Im}[\mathbb{T}_1] - \mathbb{T}_1\text{Im}[\mathbb{G}_0]\mathbb{T}_1^*]\mathbb{G}_0^*\frac{1}{1-\mathbb{T}_2^*\mathbb{G}_0^*\mathbb{T}_1^*\mathbb{G}_0^*}\right\}$$
>
> $$= \frac{2\hbar}{\pi}\int_0^\infty d\omega \frac{\omega}{e^{\frac{\hbar\omega}{k_B T_1}}-1}\text{Tr}\left\{\left[\frac{\mathcal{T}_2^\dagger + \mathcal{T}_2}{2} + \mathcal{T}_2^\dagger\mathcal{T}_2\right]\frac{1}{\mathcal{I}-\mathcal{U}\mathcal{T}_1\mathcal{U}\mathcal{T}_2}\mathcal{U}\left[\frac{\mathcal{T}_1^\dagger + \mathcal{T}_1}{2} + \mathcal{T}_1\mathcal{T}_1^\dagger\right]\frac{1}{\mathcal{I}-\mathcal{U}^\dagger\mathcal{T}_2^\dagger\mathcal{U}^\dagger\mathcal{T}_1^\dagger}\mathcal{U}^\dagger\right\}$$
>
> **Non-equilibrium force between two objects**
>
> The total force on body 2 is given by
>
> $$\mathbf{F}^{(2)} = \mathbf{F}^{(2,eq)}(T_{\text{env}}) + \sum_{\alpha=1,2}\left[\mathbf{F}_\alpha^{(2)}(T_\alpha) - \mathbf{F}_\alpha^{(2)}(T_{\text{env}})\right]$$
>
> where the first term is the usual equilibrium force (13, 14, 15, 16) and
>
> $$\mathbf{F}_1^{(2)}(T) = \frac{2\hbar}{\pi}\int_0^\infty d\omega \frac{1}{e^{\frac{\hbar\omega}{k_B T}}-1}\text{Re}\,\text{Tr}\left\{\boldsymbol{\nabla}(1+\mathbb{G}_0\mathbb{T}_2)\frac{1}{1-\mathbb{G}_0\mathbb{T}_1\mathbb{G}_0\mathbb{T}_2}\mathbb{G}_0[\text{Im}[\mathbb{T}_1] - \mathbb{T}_1\text{Im}[\mathbb{G}_0]\mathbb{T}_1^*]\mathbb{G}_0^*\frac{1}{1-\mathbb{T}_2^*\mathbb{G}_0^*\mathbb{T}_1^*\mathbb{G}_0^*}\mathbb{T}_2^*\right\}$$
>
> $$= \frac{2\hbar}{\pi}\int_0^\infty d\omega \frac{1}{e^{\frac{\hbar\omega}{k_B T}}-1}\text{Im}\,\text{Tr}\left\{\left[\mathcal{T}_2^\dagger\mathbf{p} + \mathcal{T}_2^\dagger\mathbf{p}\mathcal{T}_2\right]\frac{1}{\mathcal{I}-\mathcal{U}\mathcal{T}_1\mathcal{U}\mathcal{T}_2}\mathcal{U}\left[\frac{\mathcal{T}_1^\dagger + \mathcal{T}_1}{2} + \mathcal{T}_1\mathcal{T}_1^\dagger\right]\mathcal{U}^\dagger\frac{1}{\mathcal{I}-\mathcal{T}_2^\dagger\mathcal{U}^\dagger\mathcal{T}_1^\dagger\mathcal{U}^\dagger}\right\}$$
>
> $$\mathbf{F}_2^{(2)}(T) = \frac{2\hbar}{\pi}\int_0^\infty d\omega \frac{1}{e^{\frac{\hbar\omega}{k_B T}}-1}\text{Re}\,\text{Tr}\left\{\boldsymbol{\nabla}(1+\mathbb{G}_0\mathbb{T}_1)\frac{1}{1-\mathbb{G}_0\mathbb{T}_2\mathbb{G}_0\mathbb{T}_1}\mathbb{G}_0[\text{Im}[\mathbb{T}_2] - \mathbb{T}_2\text{Im}[\mathbb{G}_0]\mathbb{T}_2^*]\frac{1}{1-\mathbb{G}_0^*\mathbb{T}_1^*\mathbb{G}_0^*\mathbb{T}_2^*}\right\}$$
>
> $$= \frac{2\hbar}{\pi}\int_0^\infty d\omega \frac{1}{e^{\frac{\hbar\omega}{k_B T}}-1}\text{Im}\,\text{Tr}\left\{[\mathbf{p}\mathcal{U}\mathcal{T}_1\mathcal{U} + \mathbf{p}]\frac{1}{\mathcal{I}-\mathcal{T}_2\mathcal{U}\mathcal{T}_1\mathcal{U}}\left[\frac{\mathcal{T}_2^\dagger + \mathcal{T}_2}{2} + \mathcal{T}_2\mathcal{T}_2^\dagger\right]\frac{1}{\mathcal{I}-\mathcal{U}^\dagger\mathcal{T}_1^\dagger\mathcal{U}^\dagger\mathcal{T}_2^\dagger}\right\},$$
>
> where the translation operator $\mathbf{p}$ is defined by $-\boldsymbol{\nabla}_\mathbf{d}\mathcal{U}(\mathbf{d}) = \mathbf{p}\mathcal{U}$. The force on body 1 is obtained by exchanging the indices 1 and 2 in the equations above.

Instead, for a pronounced range in $R$, the emission is proportional to $R^5$, i.e., volume to the power of 5/3. While the physical origin of this dependence is nontrivial (related to the

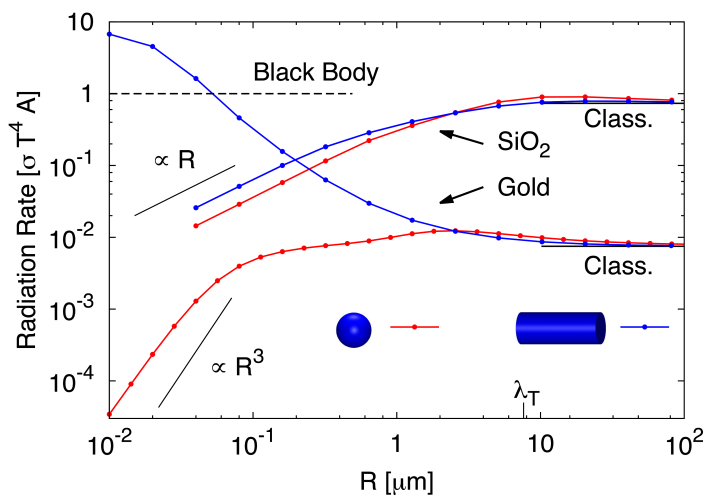

**Figure 4**

Thermal radiation of a single object at temperature $T = 300$ K as a function of radius $R$, made up of glass (SiO$_2$) or gold, as indicated. Horizontal lines on the right show the result for $R \to \infty$, given by Eq. (42) with emissivities of $\epsilon = 0.735$ for SiO$_2$ and $\epsilon = 0.0075$ for gold. Curves are normalized by the result for a black body, where $A$ is the surface area. Red curves give the radiation of a sphere, while blue curves show the equivalent results for a cylinder of infinite length. The experimental setup of the top row of Fig. 1 corresponds to $R = 0.5$ $\mu$m in the upper blue curve. The emitted radiation can be polarized, most notably for metallic materials and $R \ll \lambda_T$ (not shown) (23, 128). Data taken from Refs. (23, 128, 16).

mathematical expansion of the Mie coefficients), the physical reason for absence of volume scaling is the short skin depth $\delta$: Inside matter, the waves are damped as a function of position $x$ with $\sim e^{-x/\delta}$, so that emission is dominated by regions within a width $\delta$ from the surface. For glass, typical values for $\delta$ (which depends on wavelength) are in the micron range: Scaling proportional to volume is observed if $R$ is well below such scale. For gold, waves are damped much quicker, $\delta$ is of the order of a few nanometers, and the regime of volume-emission is shifted to much smaller values of $R$ (and not visible in the graph). We thus note that the emission of micron- or nano-scale objects is largely characterized in terms of the length scales $\lambda_T$ and $\delta$.

Turning to the second generic shape, a cylinder of infinite length, the upper blue curve of Fig. 4 shows the emission of a glass cylinder, again as a function of its radius $R$ (23, 128). The curve is very similar to that of a sphere, approaching the macroscopic law Eq. (42) for $R \gg \lambda_T$, being proportional to volume for small radii. Also the absolute values are very similar to the case of a sphere, and the mildness of optical properties of glass is manifested once more by the weak sensitivity towards shape. For gold (128), the situation is very different: While for $R \gg \lambda_T$, sphere and cylinder emit similarly, the curves are quite distinct for smaller radii. Here, the emission from the cylinder increases strongly, and even exceeds the value of unity in the units of the graph, i.e., it exceeds the emission of a black body. (Although the emissivity of any planar surface is bound from above by the value of a black body, the curve in Fig. 4 does not violate any fundamental law, as it describes a body of finite size.) This strong emissivity is remarkable, as e.g. for $R = 10$ nm, the emission of the gold cylinder exceeds the emission of a gold sphere by a factor $\sim 10^5$, so that for gold, the emitted energy depends strongly on the shape of the object.

We note that the experimental setup displayed in the top row of Fig. 1 (17) contains an $SiO_2$ nano fiber of radius 500 nm, thus found in the upper blue curve of Fig. 4. The value of the experimental $R$ is close to the regime where radiation is proportional to volume, thus distinctly different from the Stefan Boltzmann law, as observed in Ref. (17). The theoretical prediction in the graph in Fig. 1 was obtained from Eq. (41).

Last, we remark that the radiation emitted by a cylinder is in general polarized. Regarding Fig. 4, the polarization with electric fields parallel to the cylinder axis is dominant for small $R$ (for gold the degree of polarization reaches nearly 1 (128)), while for $R \approx \lambda_T$, the perpendicular component can be stronger in intensity in some cases. This effect has been observed in several experiments including Refs. (129, 22, 130).

> **Summary: Radiation of sphere and cylinder**
>
> - Eq. (42), i.e., radiation proportional to surface area, is approached only for $R \gg \lambda_T$.
> - For small $R$, $R \ll \lambda_T, \delta$, radiation is proportional to volume (not observed for gold).
> - Glass: Sphere and cylinder very similar. Gold: Sphere and cylinder are distinct.
> - Can exceed the result for a black body, as shown for a gold cylinder (nano wire).

### 5.2. Radiative Energy Transfer

Given that the radiation from an isolated object is very different from Eq. (42) when the object is small or comparable to the thermal wavelength, it stands to reason that the radiative energy transfer between two objects should also be sensitive to their separation relative to $\lambda_T$. Indeed due to *near field* effects heat transfer can be strongly enhanced (28). This is demonstrated in Fig. 5 for the case of a sphere opposite a planar surface, both made of glass ($SiO_2$) (23). The radius of the sphere is $R = 5$ μm, and we measure distance $d$ between the closest points on the surfaces. Indeed, for large distance, $d \gg \lambda_T$, the transferred energy approaches a distance independent value, as indicated by the horizontal bar. This value, although obtained for large distances, is however still a nontrivial function of radius $R$, as shown in the inset. This inset is strongly reminiscent of Fig. 4: We see a regime where the transfer is proportional to the sphere's volume, and for $R \gg \lambda_T$, $H$ becomes proportional to the surface area of the sphere, as expected. The horizontal bar in the figure, giving $H = \sigma T^4 \epsilon^2(T)$ (compare Eq. (42)), is however not approached exactly.

**Proximity approximation:** Expressing the result for curved surfacs in terms of the simpler result for parallel plates. Exact in certain limits.

For smaller distances, we see the increase of the transfer, which, in principle grows without bound. While from the experimental and technical point of view, it is also important to understand physical limitations of this growth, it is first of all important to understand its functional behavior as following from fluctuational electrodynamics (in order to detect possible deviations in experiments). Here, based on the proximity approximation (PA), one may expect that, for small $d$, $H$ diverges with $1/d$. Indeed, despite the fact that $H$ is a complicated function of the parameters involved, the following analytical expansion is valid if the sphere is small compared to the skin depth $\delta$ (131),

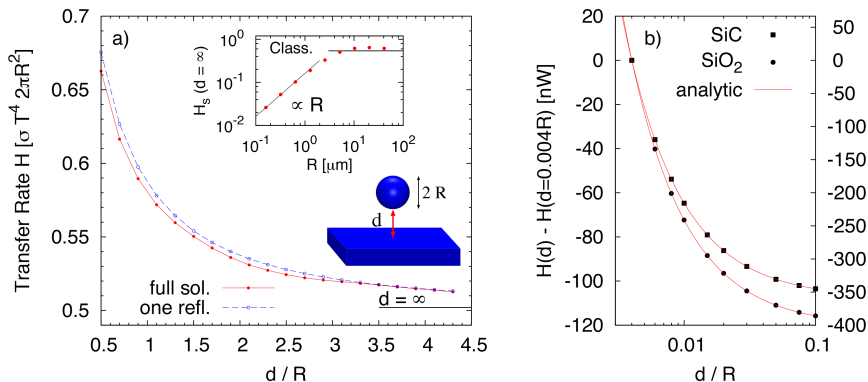

**Figure 5**

a) Heat transfer rate (in units of Stefan-Boltzmann's law) from a room temperature plate to a sphere at $T=0$ of radius $R=5\mu$m (both SiO$_2$), as a function of separation. The horizontal bar shows the limit of $d\to\infty$. The left inset shows the approach to proximity approximation for the divergent terms in a one reflection approximation. The right inset shows the result at large separation as function of $R$. b) Precision results for the same geometry, sphere and plate either both of SiO$_2$ (right abscissa) or SiC (left abscissa), as a function of distance $d$. The sphere radius is chosen very small (in the range of nanometers), so that the curve is only a function of $d/R$. Data points show numerical results, using an expansion in partial waves, and the solid lines gives the analytical result from Eq. (43). The relative deviation from numerical data to analytic curve is in the range of $10^{-4}$. Data from Refs. (23, 131).

$$H(d) = \frac{2\pi R\lambda}{d}\left[1 - (2\beta - 1)\frac{d}{R}\log\frac{d}{d_0}\right] + \mathcal{O}(d^0) \quad (43)$$

This curve is shown in Fig. 5 b) with exact numerical data, obtained by expansion in spherical waves and truncation at multipole number of 2500. The parameters $\alpha$ and $\beta$ depend on the materials and temperature, and were computed independently (not fitted) (131).

The scheme of PA (see also Refs. (42, 40, 132)), which leads in general to results similar to Eq. (43) (for cases where $R \ll \delta$ does not hold), has also proven useful in analysis of experimental data, as for example given by the dashed curve in the middle row of Fig. 1.

Figure 6 shows radiative heat transfer between a planar surface and three different generic compact objects, a sphere, a cylinder, and a cone (43). These results were obtained from a numerical scattering scheme, where the surfaces are discretized and meshed. The red curve for the sphere approaches a $1/d$ law, in agreement with Eq. (43), while the dashed lines for the cylinder asymptotically diverges with $1/d^2$. This is also in agreement with the proximity approximation, as the top of the cylinder is flat. By contrast, the transfer between the plate and the sharp cone has a much less pronounced divergence; the numerics fits consistent with a logarithmic dependence on $d$ for small $d$. This is also in qualitative agreement with the proximity approximation. In panel b), the energy flux (Poynting flux) is shown in a spatially resolved manner, as a function of lateral position on the surface. This quantity gives the relative local heating strength of the surface, which may be important for thermal writing applications. The dashed line, representing the cylinder, yields the expected result that the local heating is almost a step function with width corresponding to the diameter of the cylinder. An unexpected observation is made for the cone (which is the more dominant, the sharper the cone): The region right below the tip is actually a local minimum, while the maximal heat is absorbed a distance away from the center. This means, that the hot cone, when brought close to the surface, does not heat the surface on a localized spot, but rather on a ring around its tip: While the tip is the point closest to the surface, the cone seems to emit energy predominantly on its side, so that the competition of these effects leads to the displaced maximum.

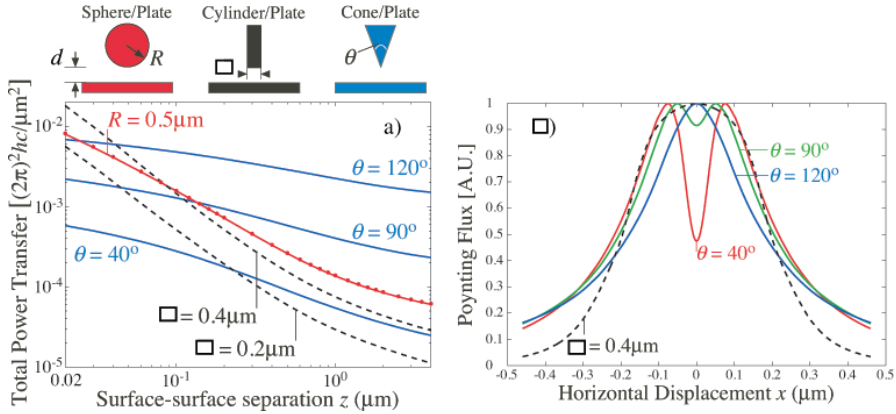

**Figure 6**
Left: Numerical results for radiative energy transfer for three different setups, as shown in the sketch, with colors of curves corresponding to colors of the sketch. Materials chosen are (marginally) different types of glass, and objects are at a temperature of 600 K, while the planar surface is kept at 300 K. Right: Spatially resolved heat transfer flux as a function of lateral position $x$ on the planar surface. Dashed curve is for the cylinder, while colored solid curves depict the cone at different opening angles. Surface-surface separation is 70 nm in all cases, and curves are normalized to a maximum of unity. From Ref. (43).

Heat transfer in the cone plate geometry has been measured in a recent experiment (36).

> **Summary: Heat transfer between sphere and plate**
>
> - For $d \gg \lambda_T$, $H \to H(R)$, where $H(R)$ has a similar form as the radiation of an isolated sphere.
>   - $H \propto R^2$ for $R \gg \lambda_T$
>   - $H \propto R^3$ for $R \ll \lambda_T, \delta$
> - For $d \ll \lambda_T, \delta$, $H \propto \frac{1}{d}$, analytic corrections are known in certain regimes.

24    Author et al.

**Summary: Heat transfer between sharp cone and plate**

- Diverges logarithmically as a function of $d$ for $d \ll \lambda_T, \delta$.
- The spatially resolved heating of the surface is maximal on a ring rather than a point.

### 5.3. Casimir Force

For systems out of equilibrium, forces acting on the objects are also quite distinct from their equilibrium counterparts. The general features are demonstrated in Fig. 7, where we show the forces on two spheres at large separation. In equilibrium, this limit is referred to as the Casimir Polder regime (133). We note the following four remarkable differences to the case of equilibrium (6):

**Force in equilibrium:**
$F^{eq} = \frac{161}{4\pi} \frac{\hbar c}{d^8} \alpha_1 \alpha_2$

**Force in equilibrium, finite temperature:**
$F^{eq} = \frac{18\hbar c}{d^7 \lambda_T} \alpha_1 \alpha_2$

1. In equilibrium, it is well known that the Casimir force between the two spheres falls off with inverse distance to the 8th power (at zero temperature) or the 7th power (at finite temperature). If out of equilibrium, i.e., if the temperatures of the two spheres and the environment are not all equal, this behavior is drastically altered, and, instead, the force decays with a much slower decay, with the inverse distance to the second power (see the mark **1** in the graph). This may be understood by contemplating the radiation pressure from photons.
2. The sign of the force, *always attractive* in equilibrium, can take either sign out of equilibrium. In the graph, we indicate repulsive forces by dashed lines, and attractive ones with solid lines. The mark **2** denotes a regime where the force is repulsive.
3. In thermal equilibrium, the Casimir force does not allow stable points, forbidden by a strict theorem. (134) Also in this respect is non-equilibrium less restrictive: The red curve in the graph changes its sign several times, so that positions of zero force exist. Naturally, in every second zero, the force changes from repulsive to attractive, such points thus exhibit stability with a minimum at the corresponding potential. (See mark **3**).
4. The graph shows the force acting on sphere 2 for different combinations of temperatures of the two spheres. We first note that the red and green curves are not identical, which directly implies that the forces on the spheres are not equal and opposite (as they would be in equilibrium). This is possible, as photons that travel to, or come from, the surrounding environment carry momentum so that the total forces on all objects in the system (including emitted and absorbed photons) still add up to zero. But staying with the example of a cold sphere (0 K) and a warm sphere (300 K) is illustrative, as then, the green curve gives the force acting on the cold one, while the red curves shows the force acting on the warm one (again they are not equal). These two curves have a crossing at roughly $d \approx 6$ $\mu$m, marked by **4**, where the forces acting on the two are equal in magnitude. The directions of forces are also equal, so that the cold sphere is repelled while the warm sphere is attracted. In this configuration, the spheres will thus start traveling in the same direction, keeping their distance constant, representing a *self propelled pair*. The effect of self propulsion through Casimir forces has been investigated in more detail in Ref. (135).

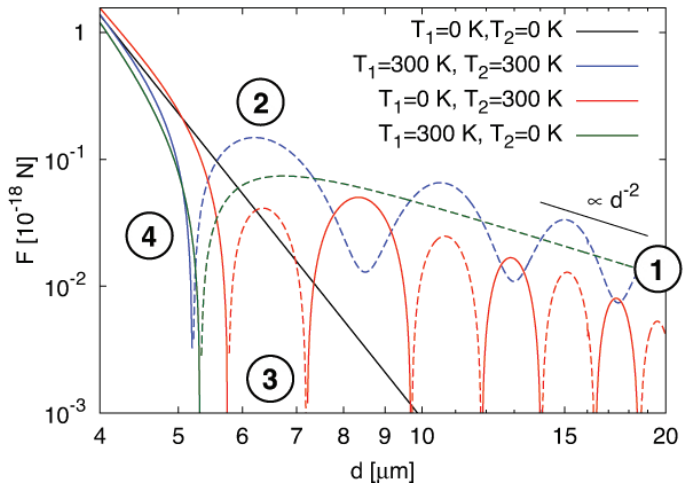

**Figure 7**
Non-equilibrium forcec between two spheres (radii $R = 1$ μm, made of SiO$_2$) at center to center distance $d$, at temperatures as given. The environment is at zero Kelvin, and thus, only the black curve describes a global equilibrium state. The curves show the force acting on the second sphere (and this force is not equal and opposite to the force acting on the first sphere). Solid lines show attraction, while dashed lines show repulsion. Four noticeable characteristics are marked with circled numbers, and discussed in the main text. From Ref. (6).

Figure 7 demonstrates the general features of non-equilibrium Casimir forces and their differences compared to their equilibrium counterparts. The experimental measurement of the forces remains however challenging, due to the fact that they mostly dominate the total force at somewhat larger distances, $d \approx \lambda_T$, and are hence often overshadowed by the equilibrium forces. In Fig. 8, we analyze a setup with potential experimental measurability (16): Here, a sphere (temperature 900 K) above surface (temperature 300 K) is considered, both of them made of a dielectric material, described by a typical Lorentz form (see Ref. (16), p. 24 for details). Here, another pronounced difference of non-equilibrium phenomena compared to equilibrium ones comes into play: In equilibrium (mathematically apparent through the evaluation on the imaginary axis), the frequency dependence of the dielectric functions enters in a broad band manner, while out of equilibrium, resonances play an important role. As such, it is for example possible to tune the non-equilibrium force from attractive to repulsive, by just tuning such resonances, especially, *their relative positions* (136, 137, 87, 16). This has been done (again, see Ref. (16), p. 24, for details), so that the sphere feels the total force depicted in the left hand side of the graph. The total force includes the gravitational force and the Casimir force which are opposite in direction. For decreasing $d$, the Casimir force, due to non-equilibrium, is repulsive, acting against gravity, and, around $\approx 0.7$ μm, balances it. For smaller $d$, the total force is repulsive. For even smaller $d$, the equilibrium part of the force will eventually take over, and render the total force attractive again. Nevertheless, at the zero force point around $d \approx 0.7$ μ, the sphere is in a local potential minimum. To make this statement even more obvious, we have in the right hand graph shown various trajectories of the sphere, if initially at $d = 800$ nm (with zero velocity): Due to the potential minimum, the sphere will perform an oscillatory motion around the potential minimum, as illustrated by the solid black curve. This curve gives however an over-idealized situation, as in reality, the sphere will cool down, due to radiative heat transfer to the surface and the environment (predominantly to the surface). This effect is included in the red curve, showing that with decreasing temperature, the repulsive force weakens, and the sphere eventually drops on the surface. Here, width of the red envelope around the curve gives the temperature difference between sphere and surface in arbitrary units. Interestingly, the relaxation time of temperature, around 1 millisecond, is on the same order of magnitude as the period of positional oscillations of the sphere. Thus, the trajectory appears experimentally measurable, and most importantly, is distinctly different from the trajectory at equilibrium, as shown by the second solid curve. Here, the Casimir force is attractive throughout, and, boosted by gravity, the sphere rapidly drops to the surface.

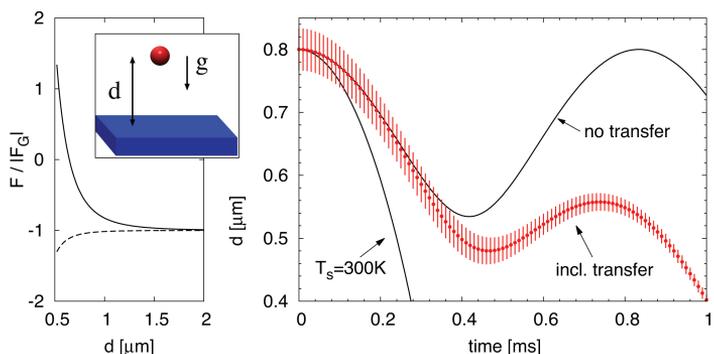

**Figure 8**
Casimir levitation: A hot sphere of $T_s = 900$ K is placed above a surface at room temperature. The left graph shows the total force, including gravity, which, at certain range of distances $d$, is repulsive. The right graph gives the trajectory of the sphere, if dropped from a height of $d = 800$ nm: The sphere bounces on the surface, a motion which is eventually cut off by the cooling of the sphere. The red curve includes this effect, and the red envelope around the curve gives the temperature difference between sphere and surface in arbitrary units. Since the trajectory is distinctly different out of equilibrium (the curve labeled $T_s = 300$ K), such a setup may provide an interesting manifestation of levitation due to out-of-equilibrium Casimir forces. Data from Ref. (16).

## 6. PERSPECTIVE

In this review we employed scattering theory along with Rytov's Fluctuational Electrodynamics to describe and gain insights into several aspects of near-field heat radiation, heat transfer, and Casimir forces. There are certainly myriad applications of these results in sub micron-scale devices, and precise experimental verification of non-equilibrium Casimir forces is lacking. There are, however, several avenues for future theoretical investigation:
• As noted in Section 2, a central assumption of Rytov's Fluctuational Electrodynam-

www.annualreviews.org • Short title    27

ics is that of local thermal equilibrium for each object. However, in real systems where temperatures of objects and environment differ substantially (e.g. hot objects in cold environment), the assumption of temperature homogeneity within each object is questionable. Hence, it would be of both theoretical and practical importance to extend the results to non-homogeneous temperatures, taking into account material heat conductivities.

• Section 5 employed bulk dielectric functions to compute heat radiation and non-equilibrium Casimir forces at scales of tens of nanometres relevant to nanofibers. To justify the approach, it is necessary to find out how accurately bulk dielectric properties apply at small scales. At scales of a few angstroms, atomistic (band structure) calculations are likely needed, while at intermediate scales a wave-number dependent dielectric response may be sufficient.

• The problem of determining the correct magnitude of the thermal contribution to the Casimir force between two metallic plates is still waiting for a definitive resolution, after twenty years of efforts. While a theoretical resolution of the conundrum might require a fully microscopic quantum mechanical theory of fluctuation forces for conduction electrons, perhaps along the lines of Ref. (75), it is hoped that an experimental clarification may be within reach based on a recently proposed differential measurement scheme (61, 62, 63, 64, 65, 66).

• The computations described here rely heavily on linear response of the bodies. It is interesting to generalize the result to the case of with non-linear electromagnetic response. Perturbative computations along this line are currently underway.

## DISCLOSURE STATEMENT

The authors are not aware of any affiliations, memberships, funding, or financial holdings that might be perceived as affecting the objectivity of this review.

## ACKNOWLEDGMENTS

MK was supported by the NSF through grant No. DMR-12-06323. MaK was supported by Deutsche Forschungsgemeinschaft (DFG) grant No. KR 3844/2-1 and MIT-Germany Seed Fund grant No. 2746830.

## APPENDIX: Heuristic Derivation of the Current Correlator

A simple derivation of Eq. 3 is possible if the dielectric is modelled as a dilute system of weakly damped isotropic harmonic oscillators. In this simple model, the electric permittivity (minus one) $\varepsilon(\omega, \mathbf{r}) - 1$ is the sum of the polarizabilities of the oscillators contained in a small volume $dV$ around $\mathbf{r}$, divided by $dV$:

$$\epsilon(\mathbf{r}, \omega) - 1 = \frac{4\pi}{dV} \sum_{j=1}^{n} \frac{e_j^2}{m_j} \frac{1}{\omega_j^2 - \omega^2 - i\omega\gamma_j}, \quad (43)$$

where $e_j, m_j, \omega_j$ and $\gamma_j$ respectively denote the charges, masses, natural frequencies and damping constants of the oscillators. The current in $dV$ is $\mathbf{I}(t) = \sum_{j=1}^{n} e_j \mathbf{v}^{(j)}(t)$, where $\mathbf{v}^{(j)}$ denotes the speed of the oscillator $o_j$. Since the speeds of distinct oscillators are uncorrelated, it is clear that the currents in non-overlapping volume elements are uncorrelated, in agreement with the $\delta(\mathbf{r} - \mathbf{r}')$ factor in Eq. 3. On the other hand, from the equations of motion of a damped harmonic oscillator one derives the following expression for correlator among the components of $\mathbf{I}$:

$$\langle I_i(t) I_k(t') \rangle = \sum_{j=1}^{n} e_j^2 \langle v_i^{(j)}(t) v_k^{(j)}(t') \rangle = \delta_{ik} \sum_{j=1}^{n} e_j^2 \langle (v_i^{(j)})^2 \rangle \cos[\omega_j'(t-t')] e^{-\gamma_j(t-t')/2}, \quad (43)$$

where $\omega_j' = \sqrt{\omega_j^2 - (\gamma_j/2)^2}$ is the (renormalized) frequency of the oscillator $o_j$, and for the last expression we note that different spatial components of the oscillators velocities are uncorrelated. In the limit of small dissipation ($\gamma_j \ll \omega_j$), taking the Fourier transform of the above correlators gives:

$$\langle I_i(\omega) I_k(\omega') \rangle = 4\pi \delta(\omega - \omega') \delta_{ik} \omega \sum_{j=1}^{n} e_j^2 \langle (v_i^{(j)})^2 \rangle \operatorname{Im}(\omega_j^2 - \omega^2 - i\omega\gamma_j)^{-1}. \quad (43)$$

In thermal equilibrium $\langle (v_j^{(j)})^2 \rangle = F(\omega_j, T)/m_j$, where $F(\omega, T) = \hbar\omega \coth[\hbar\omega/(2k_B T)]/2$ is the free energy of a (one-dimensional) quantum harmonic oscillator of frequency $\omega$ at temperature $T$. Then

$$\langle I_i(\omega) I_k(\omega') \rangle = 4\pi \delta(\omega - \omega') \delta_{ik} \omega \sum_{j=1}^{n} F(\omega_j, T) e_j^2/m_j \operatorname{Im}(\omega_j^2 - \omega^2 - i\omega\gamma_j)^{-1}$$

$$= 4\pi \delta(\omega - \omega') \omega F(\omega, T) \delta_{ik} \sum_{j=1}^{n} \frac{e_j^2}{m_j} \operatorname{Im}(\omega_j^2 - \omega^2 - i\omega\gamma_j)^{-1}, \quad (43)$$

where in the passage from the first to the second line we considered that for small dissipation all summands are significantly different from zero only for $\omega = \omega_j$ and therefore it is legitimate to replace $F(\omega_j, T)$ by $F(\omega, T)$. However $4\pi \sum_{j=1}^{n} e_j^2/m_j \operatorname{Im}(\omega_j^2 - \omega^2 - i\omega\gamma_j)^{-1} = dV \operatorname{Im} \epsilon(\omega)$, and therefore the above result is equivalent to Eq. 3 in the limit $\mathbf{r} = \mathbf{r}'$.